\documentclass[a4paper,fleqn]{cas-dc}

\usepackage{graphicx,subcaption}
\usepackage{rotating}
\usepackage{lscape}
\usepackage[justification=centering]{caption}
\usepackage{layout}
\usepackage{caption}
\usepackage[utf8]{inputenc}
\usepackage[authoryear]{natbib}

\newcommand{\Msun}{\ensuremath{M_{\odot}}}
\newcommand{\lum}{erg\,s$^{-1}$}

\newcommand{\xmm}{{\it XMM-Newton}}

\newcommand{\ergflux}{\mbox{${\rm \, erg \,\, cm^{-2} \, s^{-1}}$}}
\newcommand{\gm}{$\gamma$}

\def\tsc#1{\csdef{#1}{\textsc{\lowercase{#1}}\xspace}}
\tsc{WGM}
\tsc{QE}

\begin{document}
\UseRawInputEncoding
\let\WriteBookmarks\relax
\def\floatpagepagefraction{1}
\def\textpagefraction{.001}

\shorttitle{4FGL~J1219.0+3653}    

\shortauthors{Hazra et al.}  

\title [mode = title]{A Multiwavelength Study of the Most Distant Gamma-ray Detected BL Lacertae Object 4FGL~J1219.0+3653 ($z=3.59$)}  



%

\author[1]{Srijita Hazra}[orcid=0009-0007-0381-8069]
\affiliation[1]{organization={St. Xavier's College (Autonomous)},
            addressline={St. Xavier's College (Autonomous)}, 
            city={Mumbai},
            postcode={400001}, 
            state={Maharashtra},
            country={India}}
           
\author[2]{Vaidehi S. Paliya}[orcid=0000-0001-7774-5308]
\affiliation[2]{organization={Inter-University Center for Astronomy and Astrophysics},
            addressline={SPPU Campus}, 
            city={Pune},
            postcode={411007}, 
            state={Maharashtra},
            country={India}}
\ead{vaidehi.s.paliya@gmail.com}         
\cormark[2]
\author[3,4]{A. Dom{\'{\i}}nguez}[orcid=0000-0002-3433-4610]
\affiliation[3]{organization={Department of EMFTEL, Universidad Complutense de Madrid},
            city={Madrid},
            postcode={E-28040}, 
            country={Spain}}
\affiliation[4]{organization={Instituto de F{\'\i}sica de Particulas y del Cosmos (IPARCOS), Fac. CC. F{\'\i}sicas, Universidad Complutense de Madrid},
            city={Madrid},
            postcode={E-28040}, 
            country={Spain}}
\author[4,5]{C. Cabello}[orcid=0000-0003-4187-7055]
\affiliation[5]{organization={Departamento de F{\'\i}sica de la Tierra y Astrof{\'\i}sica, Universidad Complutense de Madrid},
            city={Madrid},
            postcode={E-28040}, 
            country={Spain}}
\author[4,5]{N. Cardiel}[orcid=0000-0002-9334-2979]
\author[4,5]{J. Gallego}[orcid=0000-0003-1439-7697]
\cortext[cor1]{Corresponding author}


\nonumnote{}

\begin{abstract}
BL Lac objects are a class of jetted active galactic nuclei that do not exhibit or have weak emission lines in their optical spectra. Recently, the first \gm-ray emitting BL Lac beyond $z=3$, 4FGL J1219.0 +3653 (hereafter J1219), was identified, i.e., within the first two billion years of the age of the universe. Here we report the results obtained from a detailed broadband study of this peculiar source by analyzing the new $\sim$58 ksec \xmm~and archival observations and reproducing the multiwavelength spectral energy distribution with the conventional one-zone leptonic radiative model. The \xmm~data revealed that J1219 is a faint X-ray emitter ($F_{\rm 0.3-10~keV}=8.39^{+4.11}_{-2.40}\times10^{-15}$ \ergflux) and exhibits a soft spectrum (0.3$-$10 keV photon index$=2.28^{+0.58}_{-0.48}$). By comparing the broadband physical properties of J1219 with $z>3$ \gm-ray detected flat spectrum radio quasars (FSRQs), we have found that it has a relatively low jet power and, similar to FSRQs, the jet power is larger than the accretion disk luminosity. We conclude that deeper multiwavelength observations will be needed to fully explore the physical properties of this unique high-redshift BL Lac object.
\end{abstract}

\begin{keywords}
 methods: data analysis \sep gamma rays: general \sep galaxies: active \sep galaxies: jets \sep BL Lacertae objects: general
\end{keywords}
\maketitle
\section{Introduction}{\label{sec:Intro}}
 Blazars are a subclass of active galactic nuclei (AGN) characterized by a prominent jet pointing within a few degrees of our line of sight (\cite{UrryandPadovani[1995]}). It is believed that synchrotron radiation from high-energy electrons spiraling along magnetic field lines within relativistic jets is the source of low-frequency emission, e.g., radio-to-optical-ultraviolet and also sometimes X-ray, from blazars. The higher energy emission detected from blazars is usually explained by the inverse Compton scattering of low-energy photons by the same population of relativistic electrons. Because of their unique orientation, several intriguing phenomena are seen in blazars caused by the relativistic amplification of the nonthermal jetted radiation \citep[see, e.g.,][]{1979rpa..book.....R}. Some examples include superluminal motion and high brightness temperature \citep{1979Natur.277..182S,2019ApJ...874...43L}, observations of temporal and spectral variability \citep{1996Natur.383..319G,2011ApJ...729....2A,2014MNRAS.441.1899F,2017ApJ...844...32P}, and detection across all accessible frequencies \citep[e.g.,][]{2011ApJ...736..131A}. The optical, radio, and X-ray emissions of blazars have also been reported to be significantly polarized \citep[e.g.,][]{2022Natur.611..677L}.  Blazars usually exhibit a flat or inverted radio spectrum. Also, because of the flux amplification, blazars are among the few types of astrophysical sources observed at cosmic distances ($z>3$) and are a dominant class of $\gamma$-ray emitters in the extragalactic high-energy sky \citep[e.g.,][]{2020ApJ...892..105A,2020ApJ...897..177P}.

 Depending on their optical spectroscopic properties, blazars have been classified as flat spectrum radio quasars (FSRQs) and BL Lac objects. FSRQs have strong, broad emission lines (rest-frame equivalent width $>$ 5\AA) in their optical spectra. In contrast, BL Lac objects show either a featureless optical continuum or an optical spectrum that displays only absorption features \citep[usually from the host galaxy, see, e.g.,][]{2011MNRAS.413..805P,2021ApJS..253...46P} or weak, narrow emission lines \citep[rest-frame equivalent width $<$ 5\AA, e.g.,][]{1991ApJ...374..431S}.
 
 Since BL Lac objects do not exhibit emission lines in their optical spectra, determining their distances (or redshifts) is a tedious task. However, the redshift measurement is essential because understanding the origins and mode of emission of relativistic jets depends on the redshift measurement. For example, the lack of the bright, high-frequency peaked BL Lacs \citep{1998MNRAS.299..433F} has been challenged by the suggestion that selection effects might be playing a role \citep[cf.][]{2012MNRAS.420.2899G}. However, as BL Lac objects pose challenges in determining their distances, the existence of the high-redshift ($z > 2$) BL Lacs may not be known even if they are present. These objects are also essential for investigating the extragalactic background light (EBL) at high redshifts \citep{2011MNRAS.410.2556D,2015ApJ...813L..34D,2018Sci...362.1031F}. Several attempts have been undertaken to find BL Lacs at high redshifts by figuring out their spectroscopic/photometric redshifts or upper/lower bounds \citep[see, e.g.,][]{2012A&A...538A..26R,2016AJ....151...35L,2018ApJ...861..130L,2017ApJ...844..120P}.

High-redshift blazars are crucial for understanding relativistic jets and their connection to the central black hole and accretion disk at the early epoch of the evolution of the universe. The supermassive black holes are reported to evolve quicker in jetted quasars compared to radio-quiet AGN, thus indicating a connection between the jet and the black hole growth \citep[e.g.,][]{2014MNRAS.442L..81F,2015MNRAS.446.2483S,2017ApJ...836L...1T}. The identification of a single blazar with bulk Lorentz factor $\Gamma$ suggests the presence of $\sim$2$\Gamma^{2}$ sources with similar intrinsic properties but having a jet pointing in a different direction \citep[cf.][]{2011MNRAS.416..216V}. Therefore, it is important to identify and study the high-redshift blazars to understand the evolution of jetted AGNs and massive black holes at the cosmic dawn \citep[see also,][]{2020ApJ...889..164M,2020MNRAS.498.2594S,2024MNRAS.528.5990S}.

4FGL~J1219.0+3653 (hereafter J1219), associated with NVSS~J121915+365718, is the first and only \gm-ray emitting BL Lac object identified above $z=3$ \citep[][]{2020ApJ...903L...8P}. Its optical and near-infrared spectra taken with 10.4 m Gran Telescopio Canarias are devoid of optical emission lines, setting a 3$\sigma$ upper limit on the accretion disk luminosity to be $2.2\times10^{44}$ \lum. The corresponding accretion rate in the Eddington unit is $<$0.2\%, assuming a black hoke mass of 10$^9$ \Msun, thus confirming the BL Lac nature of J1219 \citep[][]{2020ApJ...897..177P,2021ApJS..253...46P}. We recently obtained a $\sim$58 ksec \xmm~observations of this enigmatic blazar to unravel the physical characteristics of its relativistic jet. In this paper, we present the results of the detailed broadband study of this most distant $\gamma$-ray emitting BL Lac object carried out using \xmm~and publicly available multi-wavelength observations. We adopt a flat cosmology with $H_0 = 70~{\rm km~s^{-1}~Mpc^{-1}}$ and $\Omega_{\rm M} = 0.3$ throughout the analysis.
\section{Multi-wavelength Data Collection and Analysis}\label{sec2}
\subsection{\xmm}
\xmm~is a space-based X-ray telescope launched by the European Space Agency. It has three types of instruments onboard namely the three European Photon Imaging Cameras (EPIC), the Reflection Grating Spectrometers (RGS), and the Optical Monitor (OM) enabling simultaneous X-ray, optical and Ultra-Violet (UV) observations. The EPIC and RGS instruments work in the 0.15$-$15 keV and 0.35$-$2.5 keV energy range, respectively. On the other hand, XMM-OM is sensitive between 170 and 650 nanometres (nm). J1219 was observed by \xmm~for a net exposure of 57.9 ks (obs ID 0882470101) on 2021 November 11.

We reduce the \xmm \space data using the XMM-Newton Science Analysis Software version 21.0.0 \footnote{\hyperlink{https://www.cosmos.esa.int/web/xmm-newton}{https://www.cosmos.esa.int/web/xmm-newton}}. Since J1219 is a very faint source as revealed by previous Swift X-ray telescope observations, we verified its detection with the EPIC-PN camera using the task {\tt edetect\_chain}. The {\tt epproc} pipeline was used to produce the EPIC-PN event files which were filtered for the high-flaring background using the task {\tt evselect}. The tool {\tt evselect} was also used to extract the source spectrum from a circular region of $30^{\prime \prime}$ radius centered at the target quasar. The background spectrum was generated from a $50^{\prime \prime}$ circular region located on the same chip but free from the source contamination. The response matrix and ancillary files were created using the {\tt rmfgen} and {\tt arfgen} pipelines, respectively. The spectra were rebinned using the task {\tt specgroup} to ensure a minimum 1 count per bin. The spectral fitting was done using XSPEC \citep[][]{1996ASPC..101...17A}) and we adopted the C-statistic \citep[][]{1979ApJ...228..939C}. Given the faintness of the source, we fitted a simple power-law modified for the Galactic absorption \citep[fixed to $N_{\rm H}=1.68\times10^{20}$ cm$^{-2}$;][]{2005A&A...440..775K}. The fitting was performed in the energy range 0.3$-$10 keV. The reported errors are quoted at the 90\% confidence level.

The OM data was reduced with the task {\tt omichain}. J1219 remained undetected in all filters and 3$\sigma$ flux upper limits were estimated.

\subsection{Fermi-LAT}
The Fermi-Large Area Telescope (LAT) is a pair conversion imaging $\gamma$-ray telescope and it can scan the whole sky for an approximate period of 3.2 hours in its survey mode. It covers an energy range from 20 MeV to $>$300 GeV and has an effective area greater than 8000 $cm^2$ \citep[][]{2009ApJ...697.1071A}. More details on the LAT instrument can be found on the Fermi webpage\footnote{\url{https://fermi.gsfc.nasa.gov/ssc/}}. We collected the spectral data of J1219 from the fourth data release of the fourth Fermi-LAT \gm-ray source catalog \citep[4FGL-DR4;][]{2020ApJS..247...33A,2023arXiv230712546B}. This catalog, which contains 7194 $\gamma$-ray sources, was constructed using data from the first 14 years of the Fermi-LAT operation. This catalog covers the 50 MeV$-$1 TeV energy range and gives the sources fluxes (spectra) measured in eight energy bands: 50$-$100 MeV, 100$-$300 MeV, 300 MeV$-$1 GeV, 1$-$3 GeV, 3$-$10 GeV, 10$-$30 GeV, 30$-$100 GeV, and 100 GeV$-$1 TeV. The catalog also reports the significance of source detection in all bands. We considered those energy bands in which J1219 was detected at $\geq$2$\sigma$ confidence level. Being a faint \gm-ray emitter, the source was detected only in three energy bands.

\subsection{Archival Observations}
To collect the multi-frequency data of J1219, we utilized the Space Science Data Center (SSDC) Sky Explorer Tools \footnote{\url{https://tools.ssdc.asi.it/}} which offers flux values derived from a wide range of astronomical web databases and mission archives. Given that SSDC's source identification relies solely on the object's position in the sky, we verified that the source is included in the cited archive.

\section{Spectral Energy Distribution Modeling}\label{sec3}
We generated the broadband spectral energy distribution (SED) of J1219 using the archival and the reduced \xmm~data. We used the commonly adopted synchrotron and inverse-Compton emission model to replicate the broadband SED of J1219 \citep[e.g.,][]{2009ApJ...692...32D}. In this model, a spherical emission zone of radius $R$ is assumed to cover the whole cross-section of the jet and move down the jet with the bulk Lorentz factor $\Gamma$. In the emission zone, relativistic electrons, having a broken power law energy distribution, radiate via synchrotron, synchrotron self Compton (SSC), and external Compton (EC) processes. The magnetic field is considered to be uniform and tangled. The kinetic, magnetic, electron, and radiative jet powers were estimated following \citet[][]{30}. To compute the kinetic jet power, an equal number density of electrons and cold protons was assumed. Further details of SED modeling guidelines can be found in previous works \citep[][]{2017ApJ...851...33P,2019ApJ...881..154P,2020ApJ...897..177P}.

\begin{figure}
        \hbox{
        \includegraphics[width=\linewidth]{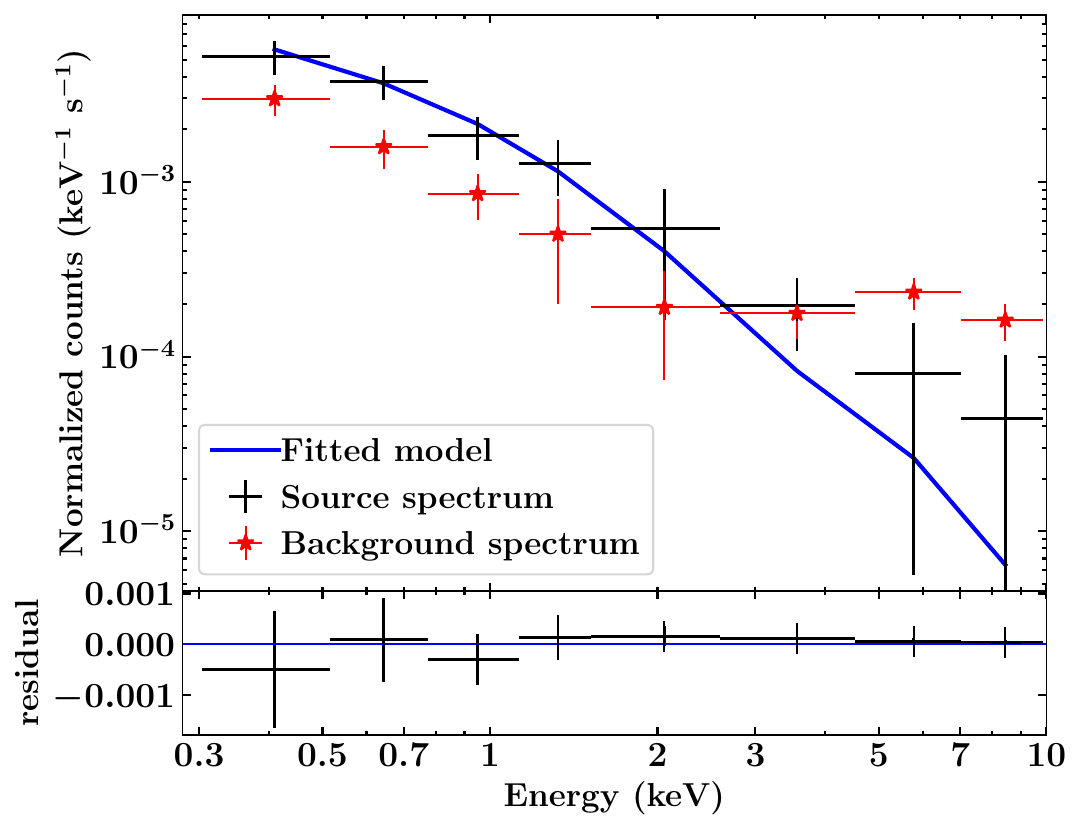}
        }
        \caption{The folded X-ray spectrum of J1219 is shown in the top panel. The black and red data points correspond to the source and background spectra, respectively. The blue line refers to the fitted absorbed power-law model. The bottom panel shows the residual of the fit.}
        \label{fig:xfit}
\end{figure}

\section{Results and Discussion}\label{sec4}
\subsection{\textit{X-ray Properties}}
The reduced \xmm~data of J1219 was fitted with a power-law model taking into account the Galactic absorption. The best-fitted X-ray photon index is $2.28^{+0.58}_{-0.46}$. The absorption-corrected 0.3$-$10 keV energy flux is $1.02^{+0.47}_{-0.24}\times10^{-14}$ \ergflux. The estimated C-statistics of the fitting is 224.42 for 231 degrees of freedom. We show the source spectrum along with the fitted power-law model in Figure~\ref{fig:xfit} and overplot the background spectrum. As can be seen, the spectrum is source dominated up to $\sim$3 keV and the background takes over at higher energies.

\begin{figure*}
        \hbox{
        \includegraphics[scale=0.235]{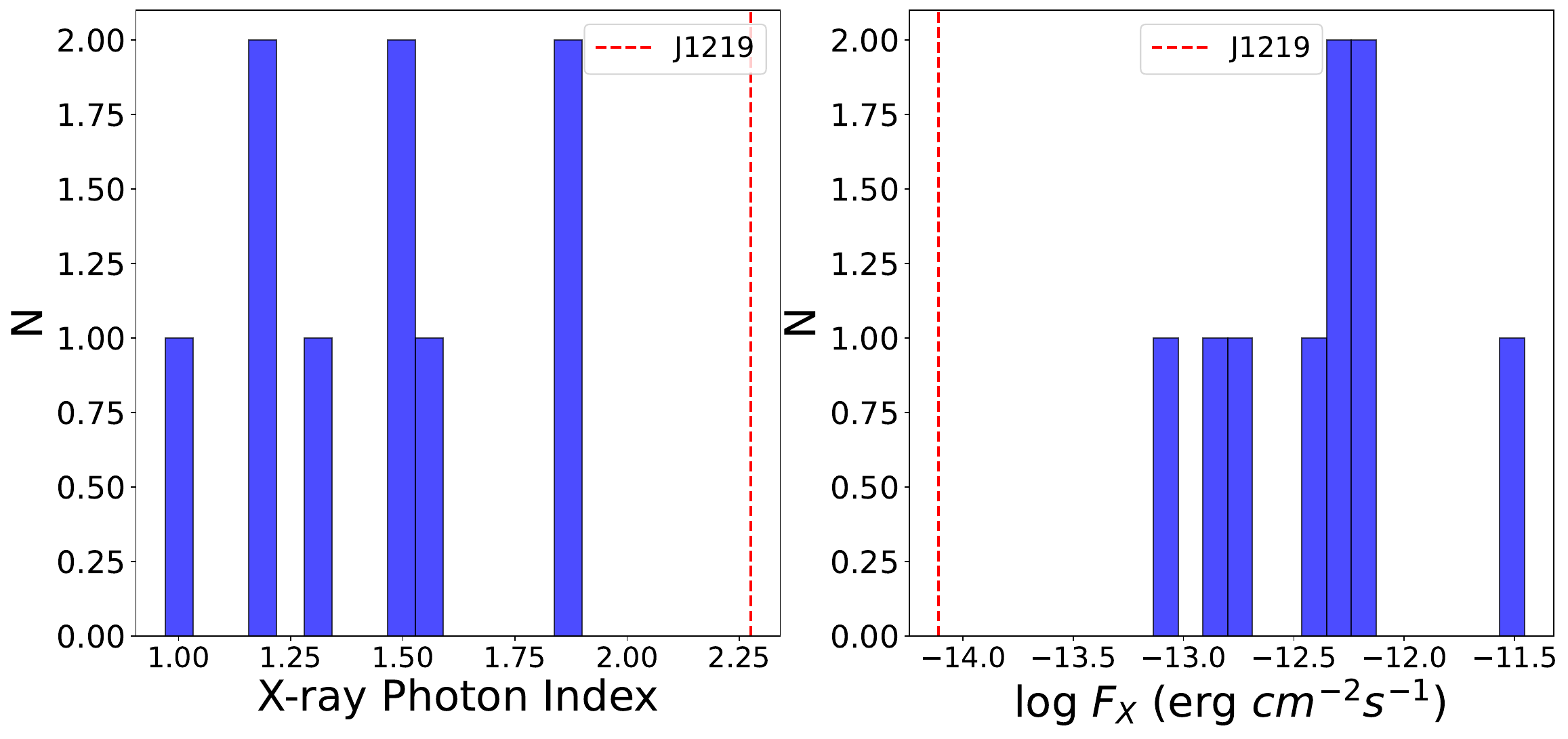}
        \includegraphics[scale=0.26]{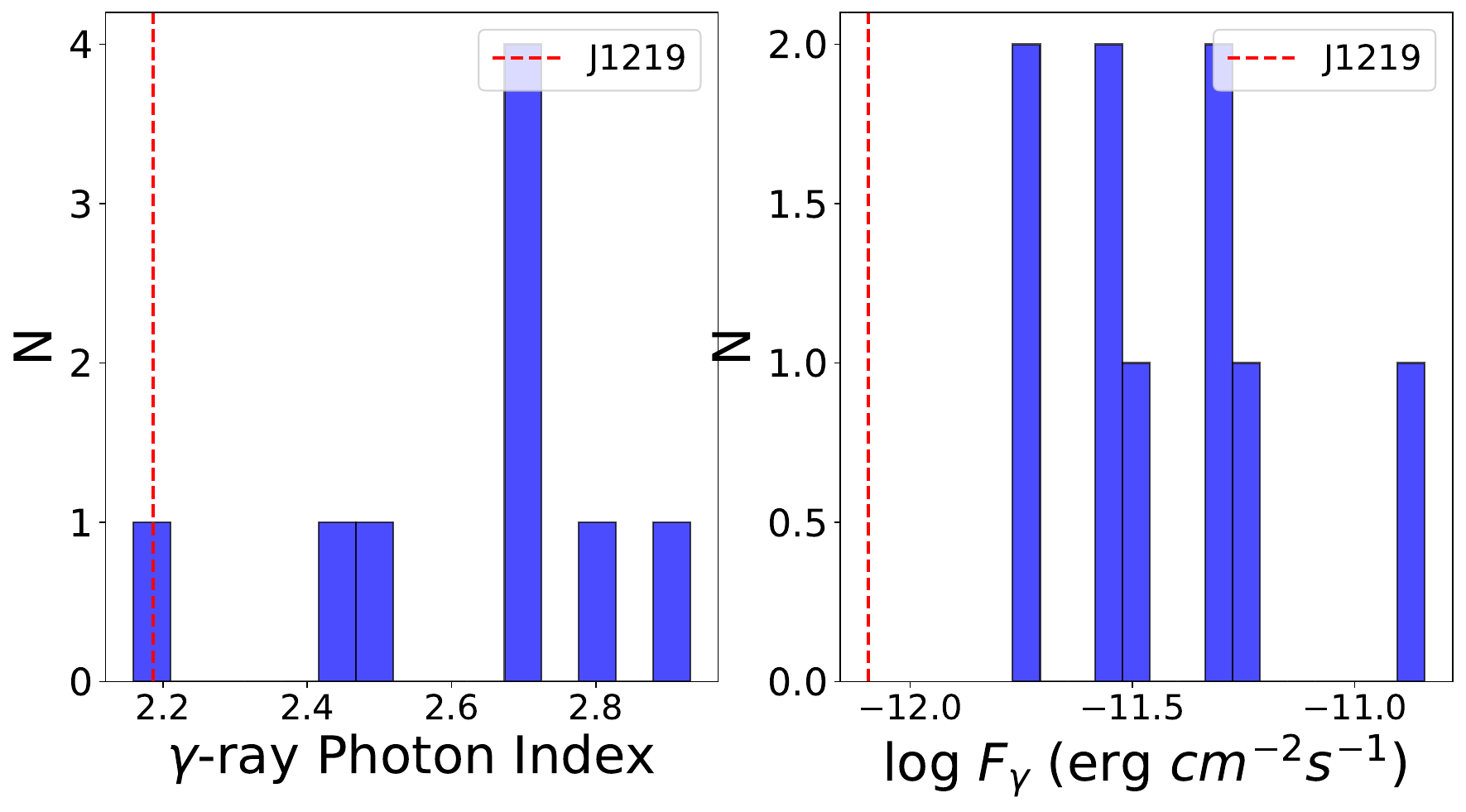}
        }
        \caption{The left two panels show histograms of the X-ray photon index and the observed X-ray flux for $z>3$ \gm-ray detected FSRQs. The right two panels refer to the \gm-ray photon index and flux. The location of J1219 is shown with the vertical dashed line.}
        \label{fig:enter-label4}
\end{figure*}

We compared the X-ray photon index and flux of J1219 with $z>3$ \gm-ray detected blazars, which are all FSRQs, studied by \citet[][]{2020ApJ...897..177P}. The results are shown in Figure~\ref{fig:enter-label4} where we highlighted J1219 by a vertical line. As can be seen, while all $z>3$ FSRQs exhibit a flat X-ray spectrum (i.e. photon index $<$ 2), J1219 has a comparatively soft spectrum. This is likely due to different radiative mechanisms responsible for the observed X-ray emission, as described in Section~\ref{subsec1}. The X-ray emission in J1219 originates from the synchrotron process. On the other hand, inverse Compton scattering mechanism is thought to be the origin of X-ray radiation in other $z>3$ blazars \citep[][]{2020ApJ...897..177P}. Furthermore, J1219 is considerably fainter in the X-ray band than $z>3$ FSRQs.

\subsection{Gamma-ray Properties}
J1219 was modeled with a log-parabola spectral model in the 4FGL-DR4 catalog. The significance of the spectral curvature over the power-law model was reported to be 2.57$\sigma$. Since the spectral curvature is not significant even at 3$\sigma$ confidence level, we preferred the results obtained from the power-law fit as given in the 4FGL-DR4 catalog. We compared the $\gamma$-ray spectral parameters, i.e., photon index and energy flux, of J1219 with $z>3$ \gm-ray emitting FSRQs \citep{2020ApJ...897..177P}. The results are shown in Figure~\ref{fig:enter-label4}. A comparison of the \gm-ray photon flux suggests that J1219 is considerably fainter than $z>3$ FSRQs. Interestingly, though the overall \gm-ray spectrum of J1219 is soft \citep[photon index$=2.25\pm0.12$;][]{2020ApJS..247...33A,2023arXiv230712546B}, it is harder than that found for $z>3$ \gm-ray detected sources. 

\begin{figure}
        \centering
        {\includegraphics[width=\linewidth]{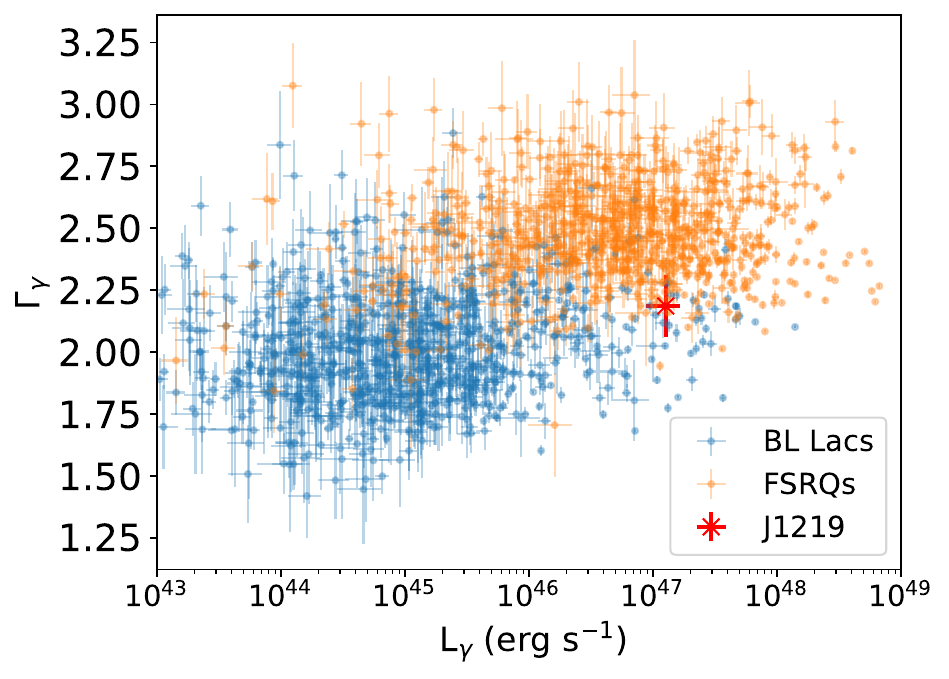}}
        \caption{Gamma-ray photon index vs gamma-ray luminosity plot for Fermi-LAT detected FSRQs (orange) and BL Lac objects (blue). The location of J1219 is shown with the red data point.}
        \label{fig:enter-label6}
\end{figure}

We highlight the location of J1219 in the \gm-ray photon index versus the luminosity plane with respect to $\gamma$-ray emitting blazars in Figure~\ref{fig:enter-label6}. Overall, J1219 has a \gm-ray luminosity similar to powerful FSRQs albeit with a harder \gm-ray spectrum. Moreover, it is among the most luminous \gm-ray detected BL Lac objects.

\subsection{Physical Characteristics Derived from the SED Modeling}\label{subsec1}
\begin{figure*}
        \centering
        {\includegraphics[width=0.85\linewidth]{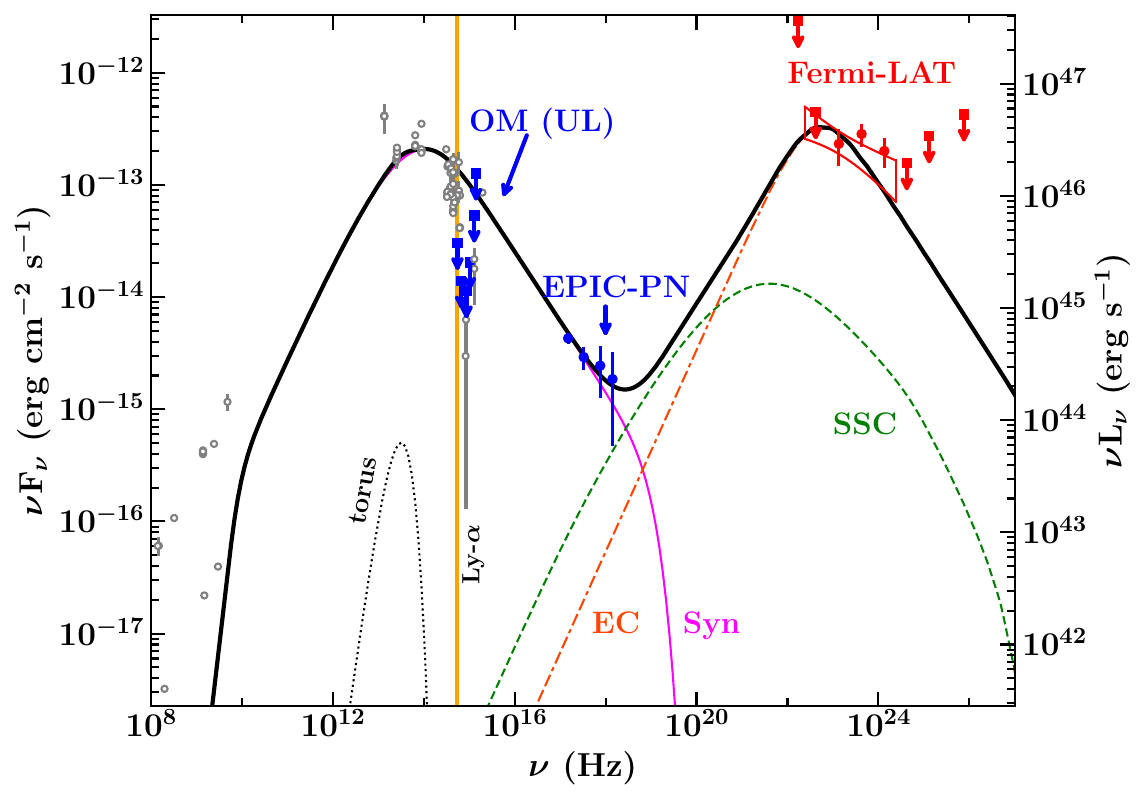}}
        \caption{The modeled SED of the farthest known $\gamma$-ray-detected BL Lac object J1219. The grey circles represent the archival data adopted from the SSDC website. The blue and red data points and arrows refer to the XMM-Newton (OM and EPIC-PN) and Fermi-LAT observations, respectively. The pink solid, green dashed, and orange dash-dash-dot lines show the synchrotron, SSC, and EC processes, respectively. The torus blackbody radiation is shown with the black dotted line. The black solid line is the sum of all the radiative components.}
        \label{fig:enter-label7}
\end{figure*}

The broadband SED of J1219 was reproduced with a single-zone leptonic emission model briefly outlined in section~\ref{sec3}. The modeled SED is shown in Figure~\ref{fig:enter-label7} and the associated SED parameters are provided in Table~\ref{tab3}. To model the radio-to-ultraviolet data with the synchrotron process, we did not consider the data points bluer than the Lyman-$\alpha$ frequency. This is because they may be affected by the absorption due to the intervening Lyman-$\alpha$ clouds whose nature is uncertain. The model fails to explain the radio observations due to the synchrotron self-absorption process.

The optical and X-ray emissions of J1219 are well explained by the synchrotron mechanism. The estimated rest-frame synchrotron peak frequency is $3\times10^{14}$ Hz which suggests the source to be an intermediate synchrotron peaked (ISP) blazar \citep[][]{2010ApJ...716...30A}. Moreover, the low X-ray flux and a soft spectrum control the SSC flux level which constrained the size of the emission region and the magnetic field. The high-energy \gm-ray emission in BL Lac objects is typically explained by the SSC mechanism, however, it was not possible to reproduce the observed \gm-ray spectrum of J1219 with it. This is because the SSC mechanism overshoots the observed X-ray flux level when trying to model the \gm-ray emission (Figure~\ref{fig:enter-label7}). A large separation between the low- and high-energy peaks also indicates that SSC alone cannot explain the high-energy emission. Therefore, we also considered the inverse Compton scattering of infrared photons produced by a putative dusty torus, i.e., the EC process. For this purpose, we assumed the torus as a spherical shell emitting blackbody radiation with an effective temperature of 1200 K and having a radius of $1\times10^{18}$ cm. 

The \gm-ray spectral curvature significance of J1219 is below 3$\sigma$ confidence level, hence not meaningful. The flux upper limits are also on the higher flux level and the inverse Compton model passes below them, thus consistent with the LAT non-detection. The three LAT detection SED points give an impression as the peak of the high-energy emission. We tried to reproduce these points with the inverse Compton peak, but failed. This is because the location of the SED peak is primarily constrained by the peak energy of the electron distribution, i.e., the break Lorentz factor ($\gamma_{\rm b}$). This parameter is tightly constrained by the optical-UV and X-ray spectral shapes, i.e., by the synchrotron spectrum. To shift the SED peak to even higher frequencies, we need to increase $\gamma_{\rm b}$, implying a higher synchrotron peak frequency; however, that conflicts with the falling X-ray spectrum and the shape of the optical-UV data points. The inverse Compton peak can also be shifted to higher frequencies by increasing the effective temperature of the considered blackbody photon field, which is dusty torus in this case. We considered it to be up to 1200 K but increasing it further may not be physically motivated since it should be below the dust sublimation temperature \citep[e.g.,][]{2011ApJ...732..116M}. Furthermore, the high-synchrotron peaked nature of J1219 is doubtful because such objects tend to have a Compton dominance (CD) less than unity \citep[cf.][]{2021ApJS..253...46P} which is against the observation (CD$\sim$1). Therefore, we also considered the \gm-ray power-law bow-tie plot during the modeling. The bow-tie plot is useful since it allows us to use the overall spectral shape in the SED modeling, especially when the target is a faint quasar, and has been used in several works \citep[][]{2010MNRAS.401.1570T}. Given the faintness of the source, the \gm-ray flux values of the data points in the individual energy bands can have some fluctuations and those may not be intrinsic to the blazar. 

The highest energy \gm-ray SED data point of J1219 reported in the 4FGL-DR4 lies in the energy range of 3$-$10 GeV. Following the EBL model of \citet[][]{2011MNRAS.410.2556D}, at $z=3.59$, the EBL optical depth for a 10 GeV photon is 0.005 and 0.80 for a 50 GeV photon. This implies a negligible EBL absorption at least up to 50 GeV. Given the falling shape of the \gm-ray spectrum, the modeled EC process predicts the \gm-ray flux level to rapidly decrease above 10 GeV. The model is free from the EBL absorption, so considering the EBL attenuation will make it even steeper. It will be possible to constrain the EBL strength when the source gets detected above 10 GeV or at even higher energies \citep[see, e.g.,][]{2024MNRAS.527.4763D}.

The 3$\sigma$ disk luminosity upper limit for J1219 ($2.2\times10^{44}$ \lum) was computed from the lack of emission lines in the optical spectrum taken with Gran Telescopio Canarias \citep[GTC;][]{2020ApJ...903L...8P}. Since the GTC spectrum was taken during a low jet activity, the disk luminosity upper limit is a fairly accurate measurement. However, given the fact that J1219 is the only known BL Lac source above $z=3$, it may not be possible to get an accurate estimate on its black hole mass ($M_{\rm bh}$). Therefore, we assumed it to be 10$^{9}$ \Msun, similar to that found for $z>3$ FSRQs \citep[][]{2020ApJ...897..177P}. The uncertainty in $M_{\rm bh}$ can primarily affect the location of the emission region since we compute the distance of the blob from the central black hole in the units of the Schwarzschild radius. For J1219, the emission region is expected to be within the torus. Since the comoving-frame radiative energy density of the torus (which depends only on the disk luminosity) is constant as long as the emission region is inside it, its exact location is unimportant for the inverse Compton calculation.

\begin{table*}
    \centering
    \caption{Parameters Used/Derived from the SED Modeling of J1219.}
    \label{tab3}
    \begin{tabular}{ll}
        \hline
    Name of the Parameter & Value \\
        \hline
        Redshift & 3.59 \\
        Bulk Lorentz factor & 15 \\
        Viewing angle & $1^\circ$ \\
        Magnetic field & 0.1 Gauss \\
        Distance of the Emission Region from the Central Black Hole & 8$\times10^{17}$ cm \\
        Radius of the emission region & 8$\times 10^{16}$ cm \\
        Dusty torus radius & 1$\times 10^{18}$ cm \\
        Comoving-frame energy density of the torus photon field & 0.04 erg cm$^{-3}$ \\
        Slope of the broken-power-law electron energy distribution before the peak & 1.4 \\
        Slope of the broken-power-law electron energy distribution after the peak & 4.25 \\
        Minimum energy of the broken power law electron energy distribution & 1 \\
        Break energy of the broken power law electron energy distribution & 3700 \\
        Maximum energy of the broken power law electron energy distribution & $10^{6}$ \\
        \hline
        Radiative jet power (\lum, log-scale) & 45.1 \\
        Kinetic jet power (\lum, log-scale) & 45.4 \\
        Electron jet power (\lum, log-scale) & 44.2 \\
        Magnetic jet power (\lum, log-scale) & 44.7 \\
        \hline
    \end{tabular}
\end{table*}

Based on the modeling, the radius of the emission region and its distance from the central black hole are $8\times10^{16}$ cm and $8\times10^{17}$ cm, respectively. The BLR or torus sizes depend on the accretion disk luminosity \citep[see, e.g.,][]{2009MNRAS.397..985G}. For a disk luminosity of $\sim2\times10^{44}$ \lum, the radii of the BLR and torus are $\sim4\times10^{16}$ cm and $\sim1\times10^{18}$ cm, respectively. This sets the emission region location outside the BLR but inside the torus. 

There are several recent works reporting the identification of so-called masquerading BL Lac objects, i.e., blazars which are intrinsically FSRQs but appear like a BL Lac due Doppler boosted jet radiation swamping out the emission lines from their optical spectra \citep[e.g.,][]{2019MNRAS.484L.104P}. The high-energy emission from some of them have also been explained with the EC mechanism \citep[cf.][]{2020ApJ...889..102R,2024ApJ...965..112R}. The GTC optical spectroscopic observation of J1219, however, was done during a low jet activity period with the continuum flux level being one third to that of the SDSS spectrum \citep[][]{2020ApJ...903L...8P}. Our SED modeling results also predict the jet to be less luminous compared to other blazars at similar redshifts, hence the jet contamination is not expected to play a major role. This suggests that the lack of the emission lines in the optical spectrum of J1219 are likely due to radiatively inefficient accretion process, rather than due to Doppler boosted jet contamination. Therefore, J1219 may not be a masquerading BL Lac. 

\subsection{Comparison with Other Fermi Blazars}\label{subsec2}
\begin{figure}
    \centering
    \includegraphics[width=\linewidth]{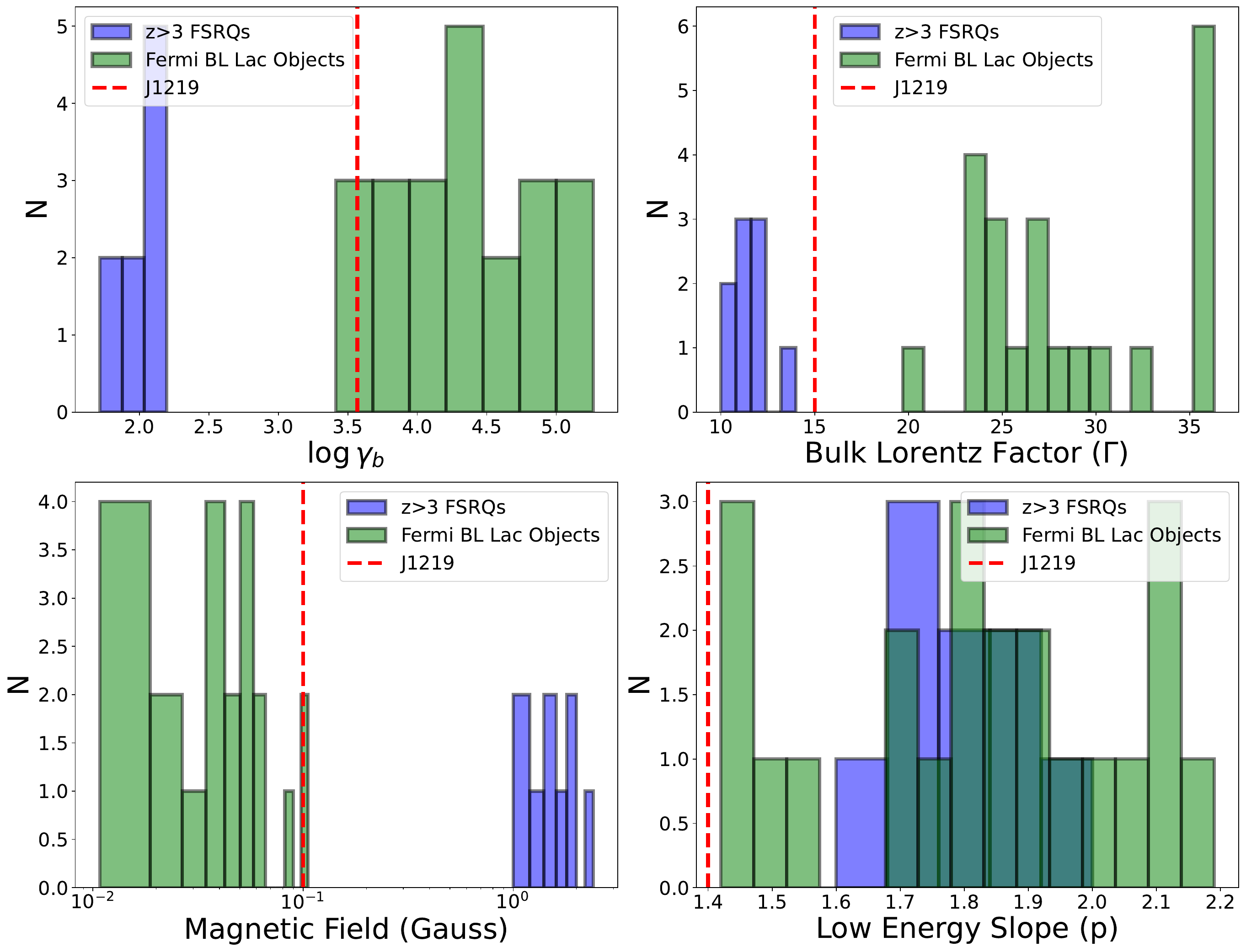}
    \includegraphics[scale=0.3]{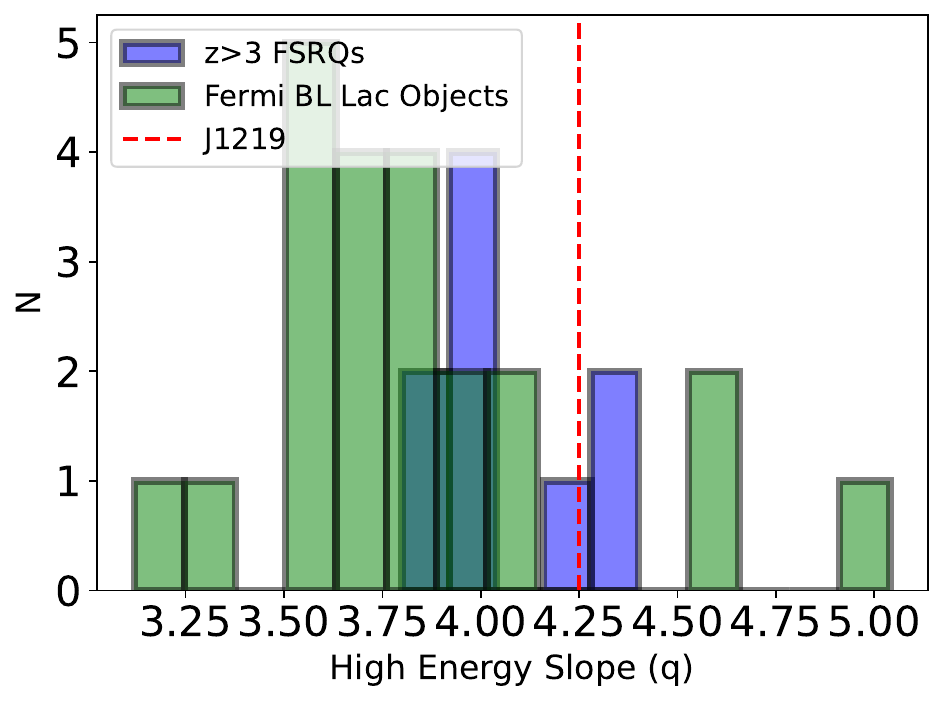}
    \caption{The histograms of various SED parameters obtained for $z>3$ FSRQs (blue) and a sample of \gm-ray emitting BL Lac objects studied by \citet[][green color]{2014MNRAS.439.2933Y}. For a comparison, we plot the location of J1219 with a vertical dashed line.}
    \label{fig:enter-label8}
\end{figure}

We briefly compared the derived SED physical parameters with that obtained by \citet[][]{2020ApJ...897..177P} for $z>3$ \gm-ray detected FSRQs. The associated histogram plots are shown in Figure~\ref{fig:enter-label8}. We also show the histograms of the SED parameters derived by \citet[][]{2014MNRAS.439.2933Y} for a sample of the Fermi-LAT detected BL Lac objects.

We found that J1219 has a bulk Lorentz factor slightly larger than $z > 3$ FSRQs and lower than other \gm-ray emitting BL Lac objects. Considering the magnetic field, it was found to be weaker than other $z>3$ blazars but similar to that derived for the BL Lac object sample. All Fermi-LAT detected $z>3$ FSRQs have a magnetic field $>$1 Gauss, whereas it was found to be 0.1 Gauss for J1219. Indeed, \gm-ray emitting BL Lac objects tend to have smaller magnetic field compared to FSRQs \citep[see, e.g.,][]{2014MNRAS.439.2933Y,2015MNRAS.448.1060G,2017ApJ...851...33P}. One possible explanation could be the fact that, in FSRQs, the \gm-ray emitting region is often predicted to lie closer to the central black hole, i.e., within the BLR or at its outer edge, where the magnetic field is higher \citep[e.g.,][]{2014Natur.515..376G,2017ApJ...851...33P}. In contrast, the emission region is possibly far from the black hole in low/intermediate synchrotron peaked (LSP/ISP) BL Lac sources where magnetic field is lower \citep[e.g.,][]{2009ApJ...707..612A}. Furthermore, the high-energy slope of the broken power law electron energy distribution of J1219 was found to be similar to $z>3$ FSRQs and other \gm-ray emitting BL Lac objects. The low-energy slope, however, appears to be on the lower side with respect to other blazars considered for comparison. In the modeling, this parameter was constrained to keep the SSC model below the observed X-ray spectrum. On the other hand, the break Lorentz factor of the particle energy distribution or $\gamma_{\rm b}$ has a considerably higher value for J1219 compared to $z>3$ FSRQs. This is because all $z>3$ FSRQs are LSP blazars, whereas, J1219, is an ISP AGN. On the other hand, this parameter is comparable to other Fermi-LAT detected BL Lac sources.

\begin{figure}
    \centering
    \includegraphics[width=\linewidth]{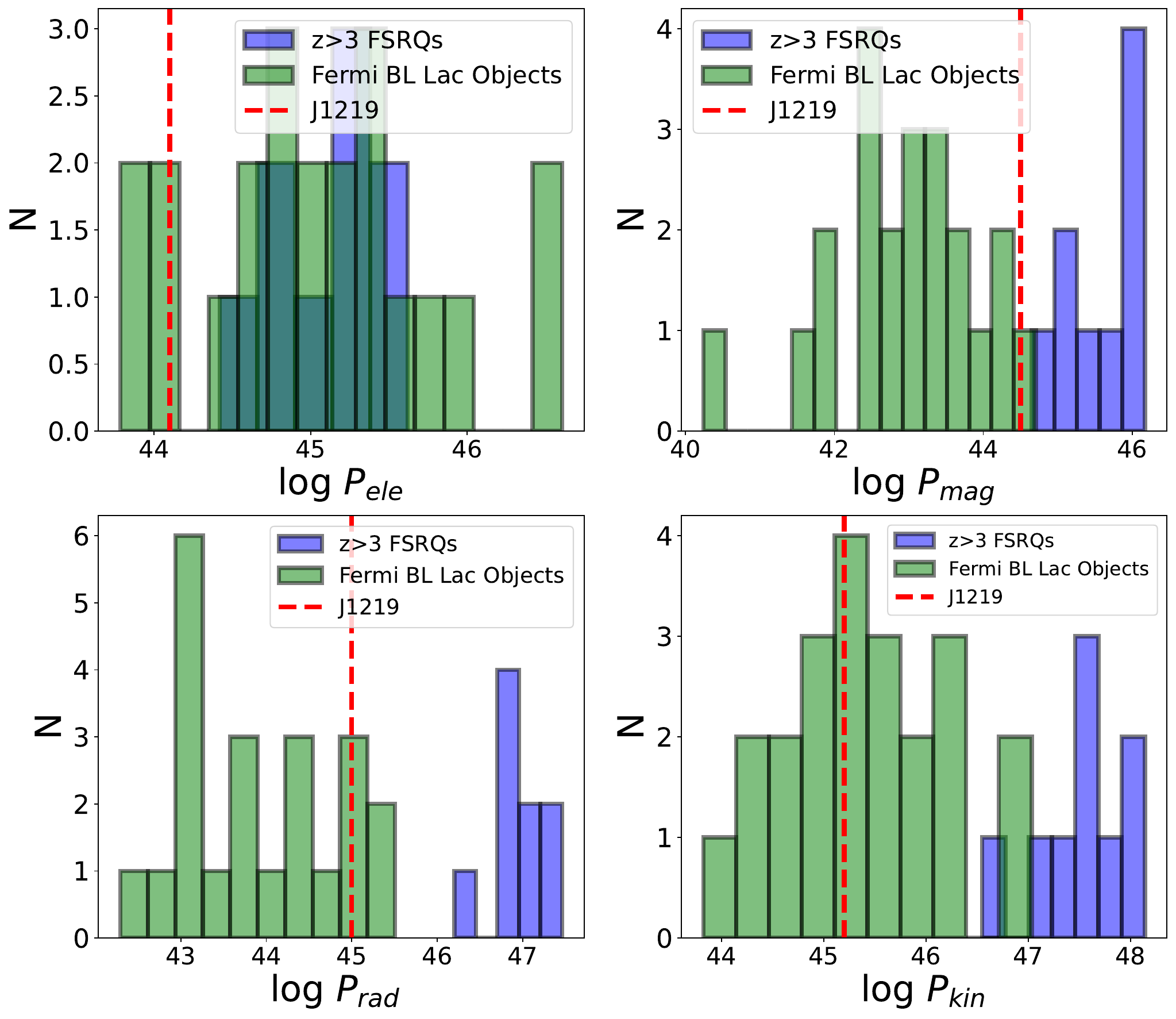}
    \caption{Same as Figure~\ref{fig:enter-label8} but for jet powers.}
    \label{fig:enter-label9}
\end{figure}

Comparing various jet powers of J1219 with $z>3$ FSRQs we found that J1219 has relatively low jet powers (Figure~\ref{fig:enter-label9}). Particularly, the radiative jet power of J1219 was found to be more than an order of magnitude lower than $z>3$ blazars. Indeed, J1219 is considerably fainter in the X- and \gm-ray bands, thus supporting the low radiative output of its jet. On the other hand, the jet powers of J1219 were found to be overall similar to other \gm-ray detected BL Lac sources. There are some minor differences, including those identified in the SED parameter comparisons (Figure~\ref{fig:enter-label8}), that can be understood by the fact that different modeling techniques were used. Since the BL Lac sample consisted of all types of blazars, i.e., LSP, ISP, and high-synchrotron peaked blazars \citep[][]{2014MNRAS.439.2933Y}, a diversity in the obtained SED parameters is expected.

\begin{figure*}
    {\includegraphics[width=\linewidth]{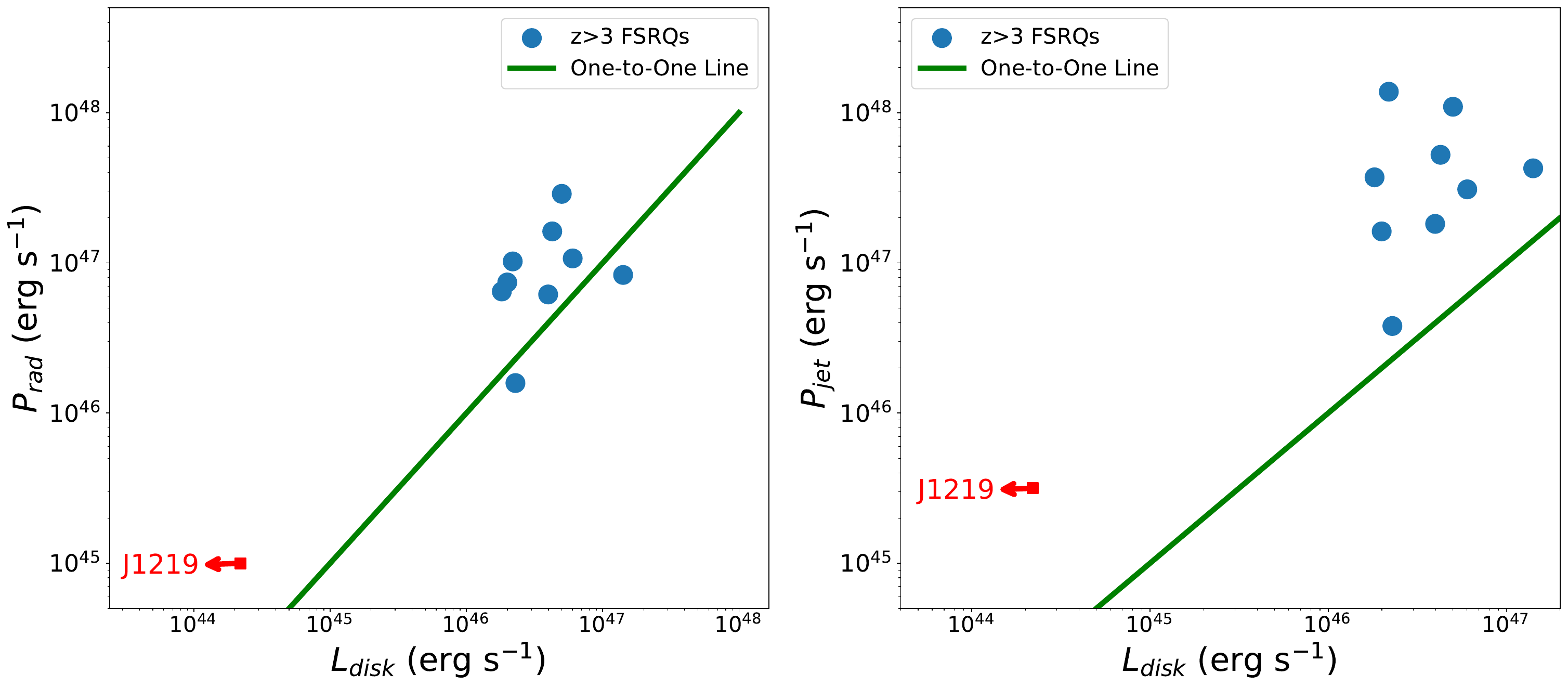}}
    \caption{Radiative jet power ($P_{rad}$) and total jet power ($P_{jet}$) as a function of $L_{disk}$ are shown in the top and bottom panels, respectively. The green line corresponds to the one-to-one correlation of the plotted quantities.}
    \label{fig:enter-label10}
\end{figure*}

Previous studies on the large samples of blazars indicated that these quantities correlate with the accretion power. The radiative jet power is found to be of the order of the disk luminosity. Moreover, The total jet power often exceeds the disk luminosity \citep[e.g.,][]{2014Natur.515..376G}. In Figure~\ref{fig:enter-label10}, we show the radiative and total jet powers as a function of the accretion disk luminosity for $z>3$ blazars. Since the accretion power is not known for J1219 due to the lack of the optical emission lines, we consider the upper limit of $2.2\times10^{44}$ \lum~estimated by \citet[][]{2020ApJ...903L...8P}.  The radiative jet power is found to be comparable to the disk luminosity for $z>3$ FSRQs; however, J1219 lies below the one-to-one correlation line. On the other hand, it lies above the one-to-one correlation line in the total jet power versus disk luminosity plot, similar to $z>3$ FSRQs.

\section{Summary}\label{sec6}
We have studied the most distant \gm-ray emitting BL Lac object J1219 using new \xmm~and other multiwavelength observations. We summarize our findings below.
\begin{enumerate}
\item J1219 is a faint X-ray emitter 9with 0.3$-$10 keV flux of $F_{\rm 0.3-10~keV}=1.02^{+0.47}_{-0.24}\times10^{-14}$ \ergflux. It exhibits a soft X-ray spectrum with 0.3$-$10 keV photon index of $2.28^{+0.58}_{-0.46}$. Comparing these spectral parameters with $z>3$ \gm-ray emitting blazars studied in \citet[][]{2020ApJ...897..177P}, we found that J1219 is the faintest X-ray source among all known $z>3$ blazars and has a steeper spectrum compared to other sources that exhibit a flat X-ray spectrum. This is likely because of the different radiation processes responsible for the observed radiation. The X-ray emission in J1219 originates from the synchrotron mechanism, whereas, it is inverse Compton process for all other $z>3$ blazars.
\item In 0.1$-$300 GeV energy range covered by the Fermi-LAT, J1219 has a considerably lower flux and relatively hard spectrum with respect to $z>3$ \gm-ray detected blazars.
\item The broadband SED of J1219 is well reproduced by a one-zone leptonic radiative model. The radio-to-X-ray spectrum of the source is explained with the synchrotron process. The estimated synchrotron peak frequency ($3\times10^{14}$ Hz) indicates it to be an ISP blazar. We found that the \gm-ray emission cannot be explained by the SSC model and EC scattering of the dusty torus can reproduce the observations. 
\item J1219 hosts a jet that has a lower radiative output compared to $z>3$ FSRQs which is also reflected in the radiative jet power versus disk luminosity plane. However, its total jet power is higher than the accretion power similar to that found for other blazars.
\end{enumerate}

Multiwavelength observations of J1219 presented in this work have provided glimpses of intriguing physical properties of this enigmatic BL Lac object. Given that it is the only known \gm-ray emitting object of this class beyond $z=3$, it will be paramount to carry out deeper observations not only to further study J1219 but also to identify more such blazars at the cosmic dawn.

\section*{Acknowledgements}
We thank the journal referee for constructive criticism. A.D. is thankful for the support of the Proyecto PID2021-126536OA-I00 funded by \break MCIN / AEI / 10.13039/501100011033. C.C. acknowledges the support of the Comunidad de Madrid for the IND2022/TIC-23643 grant. This research has made use of NASA's Astrophysics Data System Bibliographic Services. This work has also used the Space Science Data Center (SSDC), a facility of the Italian Space Agency (ASI), a multi-mission science operations, data processing, and data archiving center that supports several scientific space missions. The x-ray data is sourced from the XMM-Newton Science Archive (XSA), developed for the XMM-Newton project by the ESAC Science Data Centre (ESDC) of the European Space Agency (ESA) with requirements provided by the XMM-Newton Science Operations Centre which is located at the European Space Astronomy Centre (ESAC) in Villafranca near Madrid, Spain and responsible for the science operations of ESA's XMM-Newton satellite.
\section*{Data Availability}
All of the data used in this article lie in their respective public databases.




\bibliographystyle{elsarticle-harv} 
\bibliography{Master}



\end{document}